\documentclass[conference]{IEEEtran}

\usepackage[hidelinks]{hyperref}

\usepackage[utf8]{inputenc}
\usepackage[english]{babel}

\usepackage{bm}
\usepackage{amsthm}
\usepackage{comment}
\usepackage{xpatch}
\usepackage{color}
\usepackage{stfloats}
\usepackage[abs]{overpic}
\usepackage{cite}
\usepackage{amssymb}
\usepackage{amsmath}
\usepackage{graphicx}
\usepackage{algorithm}
\usepackage{algorithmic}
\usepackage{url} 

\usepackage{float}

\usepackage[figurename=Fig.]{caption}
\usepackage{comment}

\graphicspath{{./figures/}}

\usepackage{enumerate}

\renewcommand\IEEEkeywordsname{Index Terms}

\newcommand{\EffSNR}{\gamma_{\text{eff}}}

\newcommand{\brac}[1]{\left({#1}\right)}

\IEEEoverridecommandlockouts
\begin{document}
%
\title{Physical Layer Abstraction Model for RadioWeaves}




\author{Rimalapudi Sarvendranath, Unnikrishnan Kunnath Ganesan, Zakir Hussain Shaik, and Erik G. Larsson \\
	Department of Electrical Engineering (ISY), Link{\"o}ping University, Link{\"o}ping, SE - 581 83, Sweden.\\
	Email: \{sarvendranath.rimalapudi, unnikrishnan.kunnath.ganesan, zakir.hussain.shaik, erik.g.larsson\}@liu.se\thanks{This work was funded by the REINDEER project of the European Union‘s Horizon 2020 research and innovation program under grant agreement No.~101013425.}}

\maketitle

\begin{abstract}
RadioWeaves, in which distributed antennas with integrated radio and compute resources serve a large number of users, is envisioned to provide high data rates in next-generation wireless systems. 
In this paper, we develop a physical layer abstraction model to evaluate the performance of different RadioWeaves deployment scenarios. 
This model helps speed up system-level simulators of the RadioWeaves and is made up of two blocks.  
The first block generates a vector of signal-to-interference-plus-noise ratios (SINRs) corresponding to each coherence block, and the second block predicts the packet error rate corresponding to the SINRs generated. 
The vector of SINRs generated depends on different parameters such as the number of users, user locations,  antenna configurations, and precoders. We have also considered different antenna gain patterns, such as omni-directional and directional microstrip patch antennas.
Our model exploits the benefits of exponential effective SINR mapping (EESM).  We study the robustness and accuracy of the EESM for RadioWeaves. 
\end{abstract}

%
\IEEEpeerreviewmaketitle

\vspace{-1.2mm}
\section{Introduction}

RadioWeaves is an emerging physical-layer wireless access infrastructure, in which a fabric of distributed radio devices and computing resources serve as a massive distributed antenna array~\cite{ganesan2020radioweaves,van2019radioweaves}.
The technology builds upon the foundations of massive multiple input multiple output (MIMO) and combines the advantages of the distributed cell-free architectures and the large intelligent surfaces to achieve superior coverage at low power consumption.
RadioWeaves enables new, innovative use cases ranging from robotized factories, warehouses, immersive entertainment, and assisted living to smart homes.
RadioWeaves is foreseen to be a fundamental enabler of future 6G and beyond networks, which will offer consistent service and scalable network capacity at unprecedented energy efficiency. 

Antennas in RadioWeaves can be arranged different topologies, either in a linear topology (for instance, “radio stripe”) or in a mesh topology. In terms of spatial signal processing, RadioWeaves inherits the fundamental advantages of cellular massive MIMO: operation time-division duplexing (TDD), reliance on uplink pilots for all channel estimation tasks, and fully digital radio-frequency (RF) processing per antenna. Beyond this, there are many other benefits~\cite{van2019radioweaves}.
First, with increased spatial diversity, a terminal is likely to be close to at least a handful of antennas. This yields a superior degree of macro-diversity against signal blockage, and provides favorable path loss conditions. Second, the angular directions to the service antennas that are visible from a terminal will typically span a broad range, which causes favorable propagation conditions for the transmission of multiple data streams. 
The RadioWeaves distributed infrastructure has been shown to achieve many orders of magnitude improvements in quality-of-service and energy efficiency, compared to a conventional collocated MIMO system~\cite{ganesan2020radioweaves}. 

\begin{figure}
	\centering
	\includegraphics[width=0.7\linewidth]{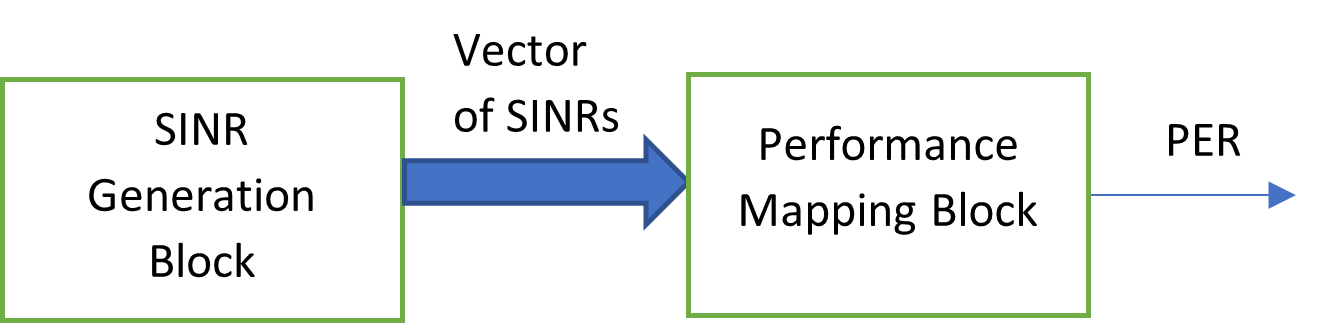}
	\caption{Physical-layer abstraction model for RadioWeaves.}
	\label{fig:pefmodel}
\end{figure}

Evaluation of the physical layer performance of this new RadioWeaves infrastructure is important. In this paper, we develop an effective signal-to-interference-plus-noise ratio (SINR) based physical-layer abstraction model to evaluate the performance of a RadioWeaves system. 
This abstraction model helps to avoid computationally expensive physical-layer link-level simulations and predicts performance with high accuracy. 
For this reason, it is routinely used for system-level simulation of many wireless  standards  such as LTE~\cite{Jiancun_2011_ICCCN,Jain_WCOM_2008},  WiMAX~\cite{tr:IEEE_EMD} and 5G-NR~\cite{Sandra_2020_ICC}. 
The physical-layer abstraction model is made up of following two blocks as shown in Fig.~\ref{fig:pefmodel}.
\begin{enumerate}
\item {\em SINR Generation Block:} The first block calculates a vector of SINRs per user for each coherence block, where the channel remains unchanged.  
\item {\em Performance Mapping Block:} The second block takes the vector of SINRs from SINR Generation Block and predicts the instantaneous packet error rate (PER) based on the modulation and coding scheme (MCS) used for the transmission.
\end{enumerate}
This performance mapping block first maps the vector of SINRs onto a scalar called the effective SINR. 
In the literature, different  mappings are considered to map vector of SINRs to effective SINR. 
For example, in {\em average value interface~\cite{tr:ESNR_FER}}, the effective SINR is the arithmetic average of the vector of SINRs. 
This is generally used to model flat-fading channels and is not an accurate measure for frequency selective channels~\cite{tr:ESNR_FER}. It is also independent of the MCS used.
In {\em capacity based mapping}~\cite{tr:ESNR_FER, Jaehyeon_VTC_2004}, the additive white Gaussian noise (AWGN) channel capacity formula is used as the mapping function.  
Here, the high SINR values impact the effective SINR more than the low SINR values as the mapping function increases monotonically. 
This makes the capacity-based mapping inaccurate as the PER is mostly impacted by the low SINR values~\cite{Jain_WCOM_2008}. 

In this paper, we study the exponential effective SINR mapping (EESM) mapping, which is obtained by  generalizing the PER of an uncoded BPSK transmission over an AWGN channel~\cite{tr:ESNR_FER,Westman_2006_MSThesis}. The generalization is done by introducing an MCS-dependent parameter. 
We focus on EESM as it has been shown to achieve a good trade-off between accuracy and computational complexity. 
It is also shown to be applicable in systems with inter-cell interference~\cite{tr:IEEE_EMD, Silva_Latincom_2011} and to OFDM systems with HARQ~\cite{Classon_2006_WCNC}. 
Furthermore, EESM has been validated to perform accurately for several wireless standards. 
It is shown to predict the performance accurately for MCS employed in IEEE 802.16 WiMAX~\cite{tr:wimax_part_16}, LTE~\cite{Jiancun_2011_ICCCN,Jain_WCOM_2008} and 5G NR~\cite{Sandra_2020_ICC}. 

In this paper, we develop an EESM based abstraction model for the RadioWeaves. 
We present the distribution of SINRs for different RadioWeaves scenarios under different precoders such as zero forcing and maximum ratio. We also compare their behavior with perfect and imperfect channel state information (CSI)  and different antenna gain patterns.
We also validate the robustness of the EESM parameter calibration for different scenarios including extreme SINR profiles.  


The following notation is used in the sequel. 
Boldface lowercase letters denote column vectors and boldface uppercase letters denote matrices. The superscripts $(\cdot)^*,~(\cdot)^\text{T},$ and $(\cdot)^\text{H}$ denote the conjugate, transpose, and Hermitian transpose, respectively. The notation $\mathbf{I}_N$ represents the $N\times N$ identity matrix. The $(m,n)$th element of a matrix $\mathbf{A}$ is denoted by $[\mathbf{A}]_{mn}$. The absolute value of a scalar and $l_2$-norm of a vector are denoted by $\vert \cdot \vert$ and $\Vert \cdot \Vert$, respectively. $\mathcal{CN}\left(\mathbf{0},\mathbf{C}\right)$ to denotes a multi-variate circularly symmetric complex Gaussian distribution with zero mean and covariance $\mathbf{C}$.

\section{System Model}
\label{sec:SystemModel}
We consider a RadioWeaves system, where $M$ antennas connected to a central processing unit (CPU) jointly serve $ K $ single antenna users. 
Let $g_{mk}\in \mathbb{C}$ denote the complex channel gain from $m$th transmit antenna to the  $k$th user located at  $ (r_{mk} , \theta_{mk} , \varphi_{mk} ) $ in the spherical coordinate system. 
Let $ \mathbf{G} = [g_{mk}]\in \mathbb{C}^{ M\times K} $ be the channel matrix between the $ M $ transmit antennas and the $ K $ users. 
We assume that the service antennas are spaced apart that coupling is insignificant. 

\subsubsection{Uplink}
\newcommand{\yul}{\mathbf{y}_{\text{ul}}}
\newcommand{\ydl}{\mathbf{y}_{\text{dl}}}

In the uplink, all the users coherently transmit data to the base station. 
The collective signal received at the service antennas, $ \yul \in \mathbb{C}^{M\times1} $, is
\begin{equation}
\yul = \sqrt{\rho_{\text{ul}}} \mathbf{G} \mathbf{D} \mathbf{q}  + \mathbf{w}_{\text{ul}},
\end{equation}
where $\rho_{\text{ul}}$ is the uplink transmit power, $\mathbf{D}$ is a diagonal matrix of power control coefficients $\text{diag}\{\sqrt{\eta_1},\ldots,\sqrt{\eta_k}\}$, $0\leq\eta_{k}\leq1$ is the power control coefficient of $k$th user, $\mathbf{q} \in \mathbb{C}^{K\times1} $ is the data from all users, and $ \mathbf{w}_{\text{ul}} \sim \mathcal{CN}(0,N_0\mathbf{I}_M) $ is the noise and $N_0$ is the noise power spectral density. 

\subsubsection{Downlink}
In the downlink, the service antennas jointly transmit the data $ \mathbf{q} \in \mathbb{C}^{K\times1} $ such that $ \mathbb{E}\{ \Vert \mathbf{q} \Vert^2 \} \leq 1$ to all the users. 
The collective signal received at the users IS denoted by the vector  $ \ydl \in \mathbb{C}^{K\times1} $ and is expressed as
\begin{equation}
\ydl = \sqrt{\rho_{\text{dl}}} \mathbf{G}^{\text{H}} \mathbf{A} \mathbf{D} \mathbf{q} + \mathbf{w}_{\text{dl}},
\end{equation}
where $\rho_{\text{dl}}$ is the downlink transmit power and $ \mathbf{A} \in \mathbb{C}^{M\times K} $ is the precoding matrix to spread the $ K $ signals into $ M $ antennas and  $ \mathbf{w}_{\text{dl}} \sim \mathcal{CN}(0,N_0\mathbf{I}_K) $ is the noise. In downlink, the power control coefficients are chosen such that $ \sum_{k=1}^{K} \eta_k \leq 1 $. 

\subsection{Channel Models}

We consider two channel models: i) Line-of-sight and ii) independent Rayleigh fading. 

\subsubsection{Line-of-Sight}

First, a line-of-sight channel for a free-space signal propagation environment is modeled. Assuming an omni-directional antenna at the user, the received power can be written as
\begin{equation}
	\label{eqn:RxPower}
	P_{rx} = \frac{P_{rad}}{4\pi} \ G(\theta,\varphi) \ \frac{1}{r^2} \ \frac{\lambda^2}{4\pi}.
\end{equation}
where
$G(\theta,\varphi)$ is the directional power gain of the transmit antenna, $P_{rad}$ is the transmitted radiated power, $r$ is the distance between the transmitter and the receiver and $\lambda$ is the wavelength of the transmitted signal.

\subsubsection{Independent Rayleigh Fading}

In a rich scattering environment, we consider the channel to be Rayleigh distributed with independent fading across antennas. Each channel gain $ g_{mk} $ is distributed as $\mathcal{CN}(0,\beta_{mk}) $, where $\beta_{mk}$ is the large scale path loss fading coefficient between transmitter and receiver with distance $d_{mk}$. It is modeled as  
$	\beta_{mk}  = -30.5 - 36.7 \log_{10} \left( {r_{mk}}/{1\text{m}}\right)$~dB~\cite{bjornson2019making}.

\subsection{Antenna Gain Pattern}
\label{ssec:AntennaDesign}

We consider two antenna gain patterns: (i) Omni-directional  and (ii) Rectangular microstrip patch antennas. Gain pattern of the omni-directional antenna is independent of the angles $\theta$ and $\varphi$, i.e., $G(\theta,\varphi) = 1$. The gain pattern of the rectangular microstrip patch antenna  is given by 
\begin{equation}
	G(\theta,\varphi) = \left( \alpha 
	\sin(\theta) \frac{\sin (X) }{X} \frac{\sin (Z)}{Z} \right)^2 ,
\end{equation}
where 
\begin{align}
	X & =\frac{\pi h}{\lambda}\sin(\theta)\cos(\varphi) , \\
	Z & =\frac{\pi W}{\lambda}\cos(\theta) , \\
	\alpha^2 & = \frac{4\pi}{\displaystyle\int_{\theta=0}^{\pi} \displaystyle\int_{\varphi=-\frac{\pi}{2}}^{\frac{\pi}{2}} \left( \frac{\sin (X)}{X} \frac{\sin (Z)}{Z} \right)^2 \sin^3(\theta) d\theta d\varphi}.
\end{align}
and  $ h $ and $ W $ denote the height and width, respectively, of the patch antenna~\cite[Ch. 14]{balanis2016antenna}. 
With vertical polarization and with no polarization losses the complex channel gain $g_{mk}$ for a microstrip patch antenna is given by \cite{balanis2016antenna},
\begin{equation}
	\label{eqn:LosChannel}
	g_{mk} = \frac{\alpha\lambda}{4\pi r_{mk}}e^{-j\frac{2\pi r_{mk}}{\lambda}}
	\sin(\theta) \frac{\sin (X) }{X} \frac{\sin (Z)}{Z}.
\end{equation}

\subsection{Channel Estimation} 

We consider least-squares (LS) channel estimation.
Let $\tau_c$ denote the length of the coherence interval \cite[Chap. 2]{marzetta2016fundamentals} and we assume there are $K$ mutually orthogonal pilot sequences of length $\tau_p$ such that $\tau_c \geq \tau_p\geq K$.
Let $\boldsymbol{\phi}_k \in\mathbb{C}^{\tau_p\times1}$ be the pilot sequence associated with user $k$, which is the $k$th column of the $\tau_p\times K$ unitary matrix $\boldsymbol{\Phi}$ and $\boldsymbol{\Phi}^\text{H}\boldsymbol{\Phi} = \mathbf{I}_K$. Each user transmits $\sqrt{\tau_p}\boldsymbol{\phi}_k^\text{H}$ over $\tau_p$ symbols and the signals collectively received  $\mathbf{Y}_p\in\mathbb{C}^{M\times \tau_p}$ can be written as 
$\mathbf{Y}_p = \sqrt{\rho_{\text{ul}}\tau_p}\mathbf{G} \boldsymbol{\Phi}^\text{H} + \mathbf{W}_p$,
where $\mathbf{W}_p$ is the noise matrix with i.i.d $\mathcal{CN}(0,N_0)$ elements. De-spreading operation is performed on the received signal $\mathbf{Y}_p$ and is given by
\begin{equation} 
	\mathbf{Y}_p' = \mathbf{Y}_p\boldsymbol{\Phi} = \sqrt{\rho_{\text{ul}}\tau_p}\mathbf{G} + \mathbf{W}_p\boldsymbol{\Phi}. 
\end{equation}
Let $\hat{g}_{mk}$ denote the LS channel estimate of $g_{mk}$. It is given by
\begin{equation}
	\label{eqn:LSEst}
	\hat{g}_{mk} = {[\mathbf{Y}_p']_{mk}}/{\sqrt{\rho_{\text{ul}}\tau_p}}.
\end{equation}
Let $\mathbf{\hat{G}} \in \mathbb{C}^{M\times K}$ denote the estimated channel matrix and let $\mathbf{\hat{g}}_k=[\hat{g}_{1k},\ldots,\hat{g}_{Mk}]^\text{T}$ is the estimated channel vector corresponding to the $k$th user.

\subsection{Precoding/Combiners} \label{sec:Precoder_Combiner}

In uplink, we consider the most commonly employed receiver combiners,  maximum-ratio combining (MRC) and zero-forcing (ZF). 
The MRC matrix is given by $\mathbf{V}^{\text{MRT}}=[\mathbf{v}^{\text{MRT}}_1,\ldots,\mathbf{v}^{\text{MRT}}_K]^\text{H}$, where  
$	\mathbf{v}^{\text{MRT}}_k = {\mathbf{\hat{g}}_k}/{\Vert \mathbf{\hat{g}}_k \Vert}$.
The ZF receiver matrix is given by
$		\mathbf{V}^{\text{ZF}} = \left(\mathbf{\hat{G}}^\text{H}\mathbf{\hat{G}}\right)^{-1}\mathbf{\hat{G}}^\text{H}$

In downlink, we consider most commonly used precoders such as maximum-ratio transmission (MRT) and ZF. 
 The MRT  matrix defined by $\mathbf{A}^{\text{MRT}}=[\mathbf{a}^{\text{MRT}}_1,\ldots,\mathbf{a}^{\text{MRT}}_K]$, where  
$		\mathbf{a}^{\text{MRT}}_k = {\mathbf{\hat{g}}_k}/{\Vert \mathbf{\hat{g}}_k \Vert}.$
The ZF precoder is given by $\mathbf{A}^{\text{ZF}} = \mathbf{\hat{G}}^\text{H}\left(\mathbf{\hat{G}}\mathbf{\hat{G}}^\text{H}\right)^{-1}$. Note that in the next section, we will use the notation $\mathbf{a}_k$ and $\mathbf{v}_k$ for downlink precoder and uplink combiner, respectively, to account for both maximum-ratio and zero-forcing.

\section{SINR Generation}\label{sec:sinr_generation}

The first block of the abstraction model shown in Fig.~\ref{fig:pefmodel} generates one SINR value per coherence block for given user based on the channel model (coherence time, coherence bandwidth)  and other parameters (carrier frequency and bandwidth).
The block also takes as input the number of users, user locations, number of antennas, antenna gain patterns, antenna configurations, channel estimator, precoder and combiner, pilot, and noise power.
Therefore, the resulting vector of SINRs depends on depends on different parameters, including large  and small scale fading.
Here, we outline the SINR computation for both uplink and downlink scenarios. 

 \newcommand{\Nr}{N_r} 
 
\subsubsection{Downlink}

In a coherence block, the collective signal received at all the users is given by 
\begin{equation}
\ydl = \sqrt{\rho_{\text{dl}}} \mathbf{G}^{\text{H}} \mathbf{A} \mathbf{D} \mathbf{q}  + \mathbf{w}. 
\end{equation}
Thus at each user, the received signal can be written as 
\begin{align}
	y_k & = \sqrt{\rho_{\text{dl}}} \mathbf{g}_k^{\text{H}} \mathbf{A} \mathbf{D} \mathbf{q}  + {w}_k \\
	& = \sqrt{\rho_{\text{dl}} \eta_k} \mathbf{g}_k^{\text{H}} \mathbf{a}_k  q_k  + \sum_{i=1,i\neq k}^K  \sqrt{\rho_{\text{dl}} \eta_i} \mathbf{g}_k^{\text{H}} \mathbf{a}_i  q_i + {w}_k.
\end{align}
The instantaneous SINR at user $ k $, is then given by
\begin{equation}
	\text{SINR}_k^{\text{inst.}} = \frac{\rho_{\text{dl}} \eta_k \vert \mathbf{g}_k^{\text{H}} \mathbf{a}_k \vert^2}{\sum_{i=1,i\neq k}^K \rho_{\text{dl}} \eta_i \vert \mathbf{g}_k^{\text{H}} \mathbf{a}_i \vert^2 + N_0},
\end{equation}
where noise power $ N_0  = k_B\cdot T\cdot BW \cdot 10^{-\text{NF}/10}$
and  $ k_B = 1.23\times 10^{-23} JK^{-1} $ is the Boltzmann constant, $ T $ (in Kelvin) is the temperature, $ BW $ (in Hertz) is the bandwidth under consideration, and $ \text{NF} $ is the noise figure in dB.
At this, we remark that for this instantaneous SINR to have a rigorous information-theoretic operational meaning, each UE needs to 
know the instantaneous value of $ \mathbf{g}_k^{\text{H}} \mathbf{a}_k$ as well as the instantaneous SINR.
These can be estimated to a good degree of accuracy by using techniques from for example \cite{interdonato2019downlink} and for the
purpose of obtaining an approximate performance prediction we will here assume that they are known.

\begin{table}[t]
	\caption{Simulation Parameters for SINR analysis}
	\label{table:SimulationSetupSINRAnalysis}
	\begin{center}
		\begin{tabular}{ |l| c| }
			\hline
			Frequency of operation, $f$ & $ 2 $~GHz  \\ \hline 
			Antenna Deployment & ULA over four walls \\ \hline
			Signal bandwidth & $20$~MHz \\ \hline
			Subcarrier bandwidth & $200$~kHz \\ \hline
			Mobility & Static \\ \hline
			Base station power & $1$~mW \\ \hline
			Base station noise figure & $ 5 $ dB \\ \hline
			User power & $1~\mu$W \\ \hline
			Pilot power & $20~\mu$W \\ \hline
			User noise figure & $ 9 $~dB \\ \hline			
			Channel type & LOS \\ \hline
			Dielectric constant, $\varepsilon_r$ & $ 10.2 $ \\ \hline
			Height of patch antenna, $h$ & $ 0.1588 $ cm \\ \hline
			Temperature of Operation & $300$~K \\ \hline
		\end{tabular}
	\end{center}
\end{table}

\begin{figure}[!htbp]
	\centering
	\includegraphics[width=0.75\columnwidth]{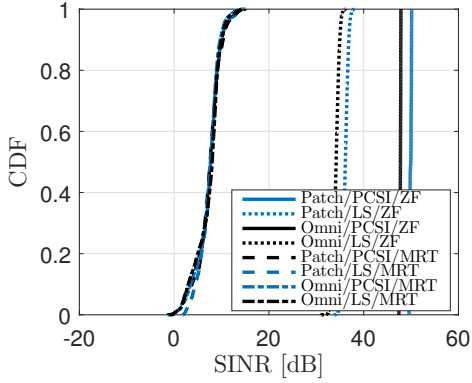}
	\caption{Downlink SINR for a room with 40mx40mx10m size, $ M=512 $, and $K=100$. The legend is to be read as antenna type/estimation method/precoder.}
	\label{fig:downlink_sinr_cdf}	
\end{figure}

\begin{figure}[!htbp]
	\centering
	\includegraphics[width=0.75\columnwidth]{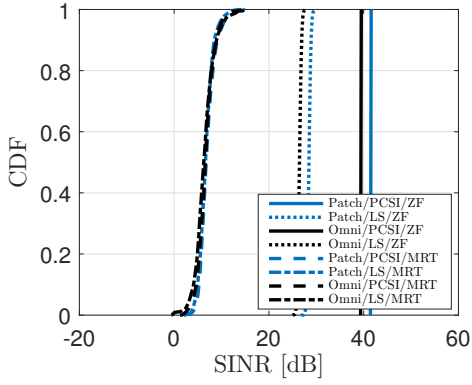}
	\caption{Downlink SINR for a room with 140mx70mx15m room size (the production hall in~\cite[p. 46]{tr:reindeer_D_1_1}), $ M=1024 $, $K=200$.}
	\label{fig:downlink_sinr_productionhall_cdf}	
\end{figure}

Figs. \ref{fig:downlink_sinr_cdf} and \ref{fig:downlink_sinr_productionhall_cdf} show the cumulative distribution function (CDF) of the downlink SINR in a RadioWeave deployment in two different  scenarios. The simulation setup is provided in Table \ref{table:SimulationSetupSINRAnalysis}. 
From the figure, it can be seen that ZF performs much better than MRT  as ZF can suppress interference from other users. 
Moreover, the ZF significantly outperforms MRT under imperfect CSI scenarios. 
The gain from the patch antennas can be exploited well by the ZF approach, while MRT is not able to exploit the   antenna gain as the interference term dominates. 
The randomness in this simulation is due to the random spatial distribution of the users, the channel estimation errors, and variations of the channel response among the subcarriers.
An important observation is that the CDFs are very stable, that is, the probability of having a scenario (in terms of coordinates
of the users) or seeing a channel realization that result in poor performance is minuscule. 




\subsubsection{Uplink}

The collective signal received from all the users at the CPU is given by 
\begin{equation}
\yul = \sqrt{\rho_{\text{ul}}} \mathbf{G} \mathbf{D} \mathbf{q}  + \mathbf{w}. 
\end{equation}
We perform a combining operation by a matrix $ \mathbf{V} \in \mathbb{C}^{K\times M} $ at the CPU, given by 
\begin{equation}
	\mathbf{V} \yul = \sqrt{\rho_{\text{ul}}} \mathbf{V}\mathbf{G} \mathbf{D} \mathbf{q}  + \mathbf{V} \mathbf{w}. 
\end{equation}
Thus the decision statistic of user $ k $ is given by
\begin{align}
	\mathbf{v}_k^\text{H}\yul & = \sqrt{\rho_{\text{ul}}} \mathbf{v}_k^{\text{H}} \mathbf{G} \mathbf{D} \mathbf{q}  + \mathbf{v}_k^{\text{H}} \mathbf{w} \\
	& = \sqrt{\rho_{\text{ul}} \eta_k} \mathbf{v}_k^{\text{H}} \mathbf{g}_k  q_k  + \sum_{i=1,i\neq k}^K  \sqrt{\rho_{\text{ul}} \eta_i} \mathbf{v}_k^{\text{H}} \mathbf{g}_i  q_i + \mathbf{v}_k^{\text{H}} \mathbf{w}.
\end{align}

As the combining vector $ \mathbf{v}_k $ is unit-norm, the instantaneous SINR at user $ k $ at the CPU is 
\begin{equation}
	\text{SINR}_k^{\text{inst.}} = \frac{\rho_{\text{ul}} \eta_k \vert \mathbf{v}_k^{\text{H}} \mathbf{g}_k \vert^2}{\sum_{i=1,i\neq k}^K \rho_{\text{ul}} \eta_i \vert \mathbf{v}_k^{\text{H}} \mathbf{g}_i \vert^2 + \sigma^2}.
\end{equation}

\begin{figure}[!t]
	\centering
	\includegraphics[width=0.75\columnwidth]{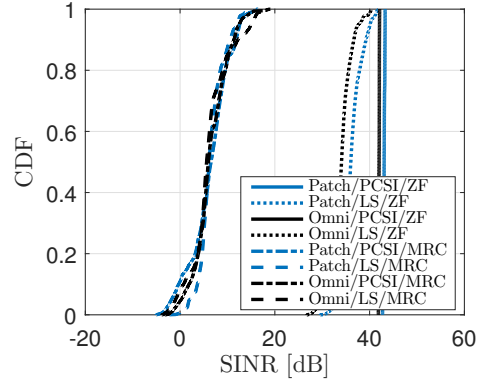}
	\caption{Uplink SINR for a room with 40mx40mx10m size, $ M=512 $, $K=100$.}
	\label{fig:uplink_sinr_cdf}	
\end{figure}

Fig. \ref{fig:uplink_sinr_cdf} shows the CDF of uplink SINR for a RadioWeave deployment. We see similar behavior as in the downlink case. For ZF, we see that the range of SINR variations similar to the downlink secnario in Fig.~\ref{fig:downlink_sinr_cdf}. However, the variation is higher for MRC.

\section{Performance Mapping}
\label{ref:Perf_mapp}

\begin{figure}
	\centering
	\includegraphics[scale=0.65]{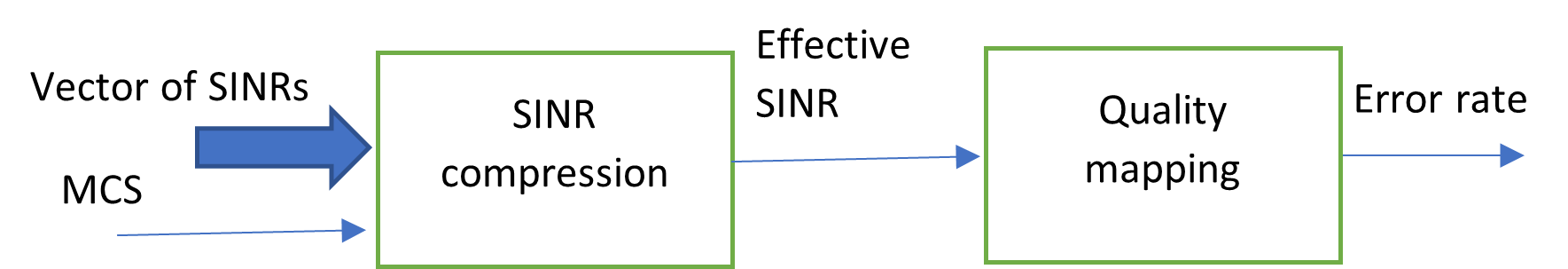}
	\caption{Performance mapping block.}
	\label{fig:mappingblock}
\end{figure}


We now describe the second block of the abstraction model.
As explained earlier, the performance mapping accurately predicts the link-level performance in a computationally efficient manner. It helps to speed up the system-level simulations as well.
The performance mapping block contains two stages, namely, \emph{SINR compression} and \emph{quality mapping} as shown in Fig.~\ref{fig:mappingblock}. The SINR compression block takes in a vector of SINRs and outputs a scalar, which is known as effective SINR. The quality mapping block maps this effective SINR to the PER. 
To understand the accuracy of the mapping, its predicted performance is compared with  actual performance obtained through link-level simulations. 
We focus on  EESM, which is shown to be accurate and has a computationally simple form.

\subsection{SINR Compression} 
The compression of the vector of SINRs onto the effective SINR combines all the different SINRs seen by the block of data into a single scalar number as each SINRs corresponds to different coherence blocks  over which data is encoded. 
Specifically, this models the link as an equivalent AWGN channel and the effective SINR can be interpreted as the equivalent signal-to-noise-ratio (SNR) on an AWGN channel. 
This compression in general depends on the MCS employed, and predicts the performance in a computationally efficient manner.

Let $\mathbf{\gamma} =[\gamma_1,\ldots,\gamma_N]^{\text{T}}$ denote the vector of SINRs and $\EffSNR$ denote the effective SINR. Then, the general form of compression is given by~\cite{tr:IEEE_EMD, tr:ESNR_FER}
\begin{equation}
	\label{Eq:Chap3_ESM}
	\EffSNR=\beta f^{-1}\brac{\frac{1}{N}\sum_{n=1}^N f\brac{\frac{\gamma_n}{\beta}}},
\end{equation}
where $f\brac{\cdot}$ is a function used to compress~\cite{tr:ESNR_FER} and $\beta$ is an MCS-dependent parameter. This parameter $\beta$ should be calibrated empirically to achieve accurate prediction.  
The accuracy of this compression is, as it turns out, very robust and a value of $\beta$ optimized for one scenario can also be
used in many other, quite different scenarios. We further elaborate on this in Section~\ref{sec:quality_mapping}. However, importantly, each MCS has a different associated $\beta$. 

For EESM,  $f\brac{\gamma_n} =1-\exp\brac{-\gamma_n}$~\cite{tr:EESM, tr:ESNR_FER, Karim_2007_MSThesis}. Therefore, the effective SINR  is given by
\begin{equation}
	\label{Eq:EESM}
	\EffSNR=-\beta \log\brac{\frac{1}{N}\sum_{n=1}^N \exp\brac{-\frac{\gamma_n}{\beta}}}.
\end{equation}
The emphasis given to lower SINR values  allows it to accurately predict the PER. 

\subsection{Quality Mapping}\label{sec:quality_mapping}
The effective SINR computed by SINR compression block is mapped to PER. 
Link-level simulations of the RadioWeaves are performed initially to calibrate this quality mapping block. Once calibration has been performed, the computationally intensive link-level simulations need not be performed again. 
For calibration of $\beta$, we have used the procedure described in~\cite[Ch. 3.2.2]{jobin_2017_phdthesis}. 
This calibration is performed only once for an MCS and the optimal $\beta^*$ is stored in a lookup table.
Table~\ref{Tbl:beta_radioWeaves} and Table~\ref{Tb2:beta_radioWeaves} show the calibrated values of $\beta^*$ for RadioWeaves for different MCS with LDPC and polar coding, respectively. With polar coding, cyclic redundancy check (CRC) bits, interleaving and rate matching are taken into consideration. 
\begin{table}[h]
	\caption{Calibrated $\beta^*$ values for RadioWeaves with LDPC codes}
	\begin{center}
		\begin{tabular}{|c | c | c | c |}
			\hline
			MCS index & Modulation & Code rate &  $\beta^*$ \\
			\hline
			0 & BPSK &1/2 & 0.78\\
			\hline
			1 & QPSK &1/2 & 1.55\\
			\hline
			2 & 16-QAM &1/2 & 4.16\\
			\hline
		\end{tabular}
	\end{center}\vspace{-4mm}
	\label{Tbl:beta_radioWeaves}
\end{table} 

\begin{table}[h]
	\caption{Calibrated $\beta^*$ values for RadioWeaves with polar codes}
	\begin{center}
		\begin{tabular}{|c| c| c| c|}
			\hline
			MCS index & Modulation & Code rate &  $\beta^*$ \\
			\hline
			1 & QPSK &1/2 & 0.624\\
			\hline
		\end{tabular}\vspace{-4mm}
	\end{center}
	\label{Tb2:beta_radioWeaves}
\end{table} 

\begin{figure}[!t]
	\centering
	\includegraphics[scale=0.35]{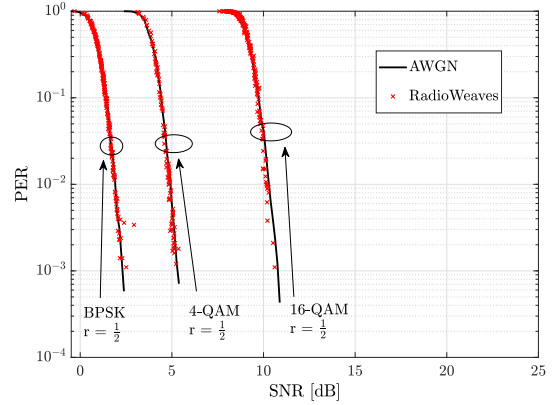}
	\caption{Validation of EESM: Performance of RadioWeaves as a function of effective SINR.}
	\label{fig:PER_Sim}
\end{figure}

Fig.~\ref{fig:PER_Sim}  plots the effective SINR $\gamma_{\text{eff}}(\beta^*)$ on the horizontal axis and the simulated PER of RadioWeaves on the  vertical axis (shown in red markers) for different realizations of the vector of SINRs. 
This is done for different MCS with LDPC codes. 
The calibrated $\beta$ values for different MCSs, which are tabulated in Table~\ref{Tbl:beta_radioWeaves}, are used in generating the plot.
We note that the markers represent the actual PER of the RadioWeaves system obtained through link-level simulations. 
They are plotted against the effective SINR obtained through the EESM mapping with calibrated $\beta^*$. 
Also, shown is the PER curve of  AWGN channel with SNR equal to the effective SINR, and for the same choices of MCS. 
We see that the markers, which represent the actual performance of the RadioWeaves system, match well with the AWGN PER values, which represent the predicted values. This establishes that the prediction accuracy of EESM is very good.

{\em Robustness of $\beta$:}
To verify the robustness of the calibration parameter $\beta$, for a given MCS, we repeated the calibration with different values of antenna elements and subcarriers.
Table~\ref{Tbl:beta_diff_params} shows the calibrated $\beta$ values with different parameters.
It clearly shows that the variation in the value of $\beta$ is negligible. 
This is in line with the conclusions in~\cite{Sandra_2020_ICC}, where the variation in $\beta^{*}$ values is shown to be negligible as the subcarrier spacing or other 5G-NR numerology changes.

\begin{table}
	\caption{Calibrated $\beta^*$ values for different parameter values (MCS index =1)}
	\begin{center}
		\begin{tabular}{ |c |c| c|}
 \hline
 Number of antennas & Number of subcarriers & $\beta^*$ \\
 \hline
 48                 & 2                     & 1.54      \\
 \hline
 48                 & 18                    & 1.55      \\
 \hline
 64                 & 2                     & 1.54      \\
 \hline
 64                 & 18                    & 1.55      \\
 \hline		\end{tabular}
	\end{center}\vspace{-8mm}
	\label{Tbl:beta_diff_params}
\end{table} 

Furthermore, in Table~\ref{Tbl:per_diff_beta}, we compared the simulated PER and the predicted PER for different parameters values for which calibration is not performed. 
Here, the simulated PER is obtained through the RadioWeaves link-level simulations and  the predicted PER is obtained by the EESM-based performance mapping.
It also compares the simulated PER with predicted PER for two values of $\beta^{*}$
from Table~\ref{Tbl:beta_diff_params}. We see that the predicted values are very close to the simulated and they change only by a small value as $\beta^{*}$ varies slightly.

\begin{table}
	\caption{Comparison of simulated PER with predicted PER for different values of $\beta^*$ (M=72, QPSK, and rate=0.5)}
	\begin{center}
		\begin{tabular}{|c|c|c|}
			\hline
			Simulated PER & Predicted PER ($\beta^{*}$=1.54) & Predicted PER ($\beta^{*}$=1.6) \\  \hline
			0.0027     &              0.0039              & 0.0038                          \\  \hline
			0.0015     &              0.0009              & 0.0009                          \\  \hline
			0.0008     &              0.0006              & 0.0005                          \\  \hline
			0.5040     &              0.4849              & 0.4830                          \\  \hline
			0.2920     &              0.3081              & 0.3059                          \\  \hline
			0.2760     &              0.2714              & 0.2693  						    \\  \hline
		\end{tabular} 
	\end{center}\vspace{-2mm}
	\label{Tbl:per_diff_beta}
\end{table} 

\begin{figure}[!t]
	\centering
	\includegraphics[scale=0.35]{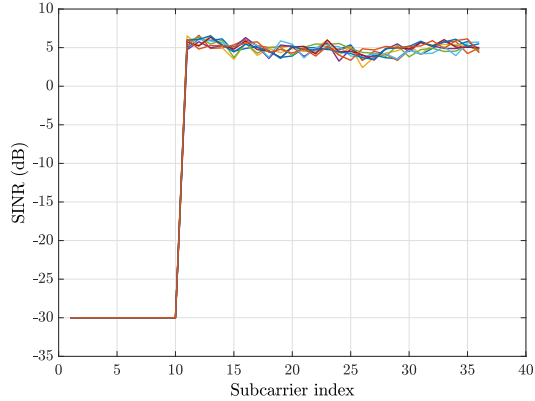}
\caption{Extreme SINR profile as a function of subcarrier index. Here, different curves correspond to different realizations for which PER's are shown in Table~\ref{Tbl:extrem_sinr}.  }
	\label{fig:extremesnrcheck}
\end{figure}

{\em Extreme SINR profile:} We now consider an extreme, artificial example in order to demonstrate the robustness of the EESM technique. We specifically study the behavior of the performance mapping  for an  SINR profile that is not used for calibration of $\beta$. This SINR profile has a set of  SINR values that  are extremely low, as shown in Fig.~\ref{fig:extremesnrcheck}.
Table~\ref{Tbl:extrem_sinr} compares the simulated PER with the predicted PER for $\beta^{*}=1.54$.
We see that the predicted PERs are reasonably close to the simulated PER values even for this extreme SINR profile for which calibration is not done.

\begin{table}
	\caption{Comparison of simulated PER with predicted PER for extreme SINR profile shown in Fig.~\ref{fig:extremesnrcheck} for M=72, QPSK, and rate=0.5. }
	\begin{center}
		\begin{tabular}{|c|c|}
			\hline
			Simulated PER & Predicted PER ($\beta^{*}$=1.54) \\ \hline
			0.0187     &    0.0173     \\ \hline
			0.0251     &    0.0254     \\ \hline
			0.0256     &    0.0199     \\ \hline
			0.0222     &    0.0188     \\ \hline6
			0.0255     &    0.0212     \\ \hline
			0.0145     &    0.0163     \\ \hline
			0.0138     &    0.0156     \\ \hline
			0.0118     &    0.0162     \\ \hline
		\end{tabular} \vspace{-4mm}
	\end{center}
	\label{Tbl:extrem_sinr}
\end{table} 

\section{Conclusion}\label{sec:conclusion}

We presented a physical-layer abstraction model for the RadioWeaves infrastructure.
This model enables the prediction of the RadioWeaves physical layer performance in a computationally efficient manner with high accuracy. 
It can especially be used to speed up system-level simulations of RadioWeaves. 
Our simulations showed that EESM performs well for different MCSs.
The study has also shown that the calibration parameter is mainly dependent on the MCS used and that once it is obtained for a particular MCS, it can be used across a wide range of scenarios with different parameters such as the number of antennas and the number of subcarriers. 
We showed that the EESM is a highly robust way of mapping a vector of instantaneous SINRs to link-level performance in terms of PERs for a variety of scenarios, where calibration is required only once and where the specific scenario selected for the calibration is not particularly important.

\bibliographystyle{IEEEtran}
\bibliography{IEEEabrv,references}

\end{document}